\documentclass[aps, prd, amsmath, floats,floatfix, twocolumn, superscriptaddress,nofootinbib, showpacs]{revtex4}
 
\usepackage{amssymb}
\usepackage{amsmath}
\usepackage{verbatim}
\usepackage{mathrsfs}
\usepackage{amsfonts}
\usepackage{latexsym}
\usepackage{epsfig}
\usepackage{color}
\usepackage{graphicx,subfigure}
\usepackage{units}
\usepackage{slashbox}

\begin{document}

\title{No-go theorem for static scalar field dark matter halos with no Noether charges}

\author{Alberto Diez-Tejedor}
\affiliation{Departamento de F\'isica, Divisi\'on de Ciencias e Ingenier\'ias, Campus Le\'on, 
Universidad de Guanajuato, Le\'on 37150, M\'exico}

\author{Alma X. Gonzalez-Morales}
\affiliation{Departamento de F\'isica, Divisi\'on de Ciencias e Ingenier\'ias, Campus Le\'on, 
Universidad de Guanajuato, Le\'on 37150, M\'exico}
\affiliation{Instituto de Ciencias Nucleares, Universidad Nacional
  Aut\'onoma de M\'exico, Circuito Exterior, C.U., A.P. 70-543,
  M\'exico D.F. 04510, M\'exico}

\date{\today}

\begin{abstract} 
Classical scalar fields have been considered as a possible effective description of dark matter. We show that, for any metric theory of gravity, no static, spherically symmetric, regular, spatially localized, attractive, stable spacetime configuration can be sourced by the coherent excitation of a scalar field with positive definite energy density and no Noether charges. In the weak-field regime the result also applies for configurations with a repulsive gravitational potential. This extends Derrick's theorem to the case of a general (non-canonical) scalar field, including the self-gravitational effects. Some possible ways out are briefly discussed.
\end{abstract}

\pacs{
95.35.+d, 
04.40.-b, 
03.50.-z 
}

\maketitle

\section{Introduction}

There is not yet a definite answer to the dark matter (DM) problem. At the fundamental level, DM should probably be described in terms of a quantum field theory. There has been much progress in this direction within the last few decades~\cite{Bertone:2004pz}, although direct~\cite{direct} and indirect~\cite{indirect} detection methods are still inconclusive. From a different perspective, it would be possible that, at the effective level, DM admits a classical description, e.g. if the DM particles develop a Bose-Einstein condensate~\cite{BEC}, or reach a hydrodynamic regime~\cite{hydro}. In this paper we will consider that a metric theory (not necessarily Einstein) describes the gravitational interaction, and restrict our attention to classical scalar field theories.

As candidates to describe the DM, scalar fields are expected to develop static, spherically symmetric, regular, spatially localized, attractive, stable, self-gravitating spacetime configurations, that can be identified with galactic halos. We show that, with no global symmetries in the action, these configurations are only possible at the expense of having negative energy densities.

In flat spacetime and for the case of a canonical scalar field this is a consequence of  Derrick's theorem~\cite{derrick}: in three spatial dimensions there are no regular, static, localized scalar field configurations with positive definite energy density (today we know several means, either topological~\cite{topological} or non-topological~\cite{non-topological}, to evade this theorem). Here we extend this result to the case of a general scalar field with an arbitrary kinetic term, including the self-gravitational effects. 

This is not only a purely academic exercise. In flat spacetime a static, spherically symmetric, spatially localized perfect fluid distribution is necessarily trivial, but non-trivial self-gravitating solutions do exist~\cite{Tolman:1939jz}. These solutions have been used thoroughly for the study of stellar structure~\cite{ns}. Nothing prevents something similar from happening for scalar field configurations; it is then natural to look for a version of Derrick's theorem in presence of gravity. Real galaxies may not match all the conditions in the theorem, however, large deviations are not expected: presumably DM halos have a small angular momentum~\cite{a.momentum} and triaxiality~\cite{triaxiality.1}; see Ref.~\cite{triaxiality.2} for a debate. The existence of a smooth transformation from idealized halos to actual ones gives some (astro)physical support to the results in this paper. 

To proceed we will consider the most general action that can be constructed from a real, minimally coupled scalar field and its first derivatives, 
\begin{equation}
 S =\int d^4 x \sqrt{-g} \, M^4\mathcal{L}(\phi/M, X/M^4)\,. \label{action}
\end{equation}
We assume that the theory is local, and Lorentz invariant. Here $\mathcal{L}$ is the Lagrangian density, and $X\equiv - \frac{1}{2}\partial_{\mu}\phi \partial^{\mu}\phi$ the kinetic scalar. The coupling of this field to the standard model of particle physics is highly constrained by observations, and in this paper it is considered to be negligible. We are adopting the mostly plus signature $(-,+,+,+)$ for the spacetime metric, and taking units with $4\pi G=c=\hbar=1$. The characteristic scale $M$ and the scalar field $\phi$ are measured in units of energy. A theory of the form in Eq.~(\ref{action}) is appropriate for the description of a single scalar degree of freedom, and is usually dubbed {\it k-essence}~\cite{ArmendarizPicon:1999rj}. We will discuss the case with more than one field at the end of the paper.

With the notation in Eq.~(\ref{action}), the Lagrangian density for a canonical scalar field takes the form $\mathcal{L}_{\textrm{can}}=X-M^4V(\phi/M)$, with $M^4 V(\phi/M)$ a potential term~\cite{Turner:1983he}. 
If the Lagrangian density depends only on the kinetic scalar the resulting theory is called purely-kinetic \cite{Scherrer:2004au}. 

In order to have a sensible theory, the Hamiltonian should be bounded from below. In particular, we will adopt the weak energy condition; that is, for every future-pointing timelike vector field $t^{\mu}$, the energy density measured by the corresponding observers should always be non-negative, $\rho_t\equiv T_{\mu\nu}t^{\mu}t^{\nu}\ge 0$. Otherwise, a vacuum energy scale would appear in the theory, bringing back fine-tuning issues usually associated with the cosmological constant problem.

We also neglect higher-derivative terms in Eq.~(\ref{action}): On the one hand, they source extra dynamical degrees of freedom, most of which are not $-$generically$-$ well behaved; see however Ref.~\cite{Nicolis:2008in}. Additionally, these new degrees of freedom couple gravitationally to the standard matter, introducing departures from general relativity. (If dark matter exists, general relativity would probably describe the gravitational interaction at galactic scales.)

\section{Static scalar field configurations}\label{sec.static}

The behavior of a scalar field depends crucially on the character of the derivative terms. If they are timelike, $X>0$, the energy-momentum tensor can be formally identified with that of a perfect fluid, i.e. $p_{\|}=p_{\perp}$ in Eq.~(\ref{energymomentum}) below.  On the contrary, if the derivative terms are space-like, $X<0$, the energy-momentum tensor of the scalar field takes the form of a relativistic anisotropic fluid with 
\begin{equation}\label{relani}
 p_{\perp}=-\rho=\mathcal{L} \, , \quad 
 p_{\|}=\mathcal{L}-2X\partial\mathcal{L}/\partial X \, .
\end{equation}
Here $\rho$ is a energy density, and $p_{\|}$ and $p_{\perp}$ are a longitudinal and a transverse pressures, respectively. From now on and in order to simplify the notation we will omit the characteristic scale $M$. 

In cosmology, the homogeneous and isotropic background guarantees a perfect fluid description.  However, static spacetime configurations restrict the possible scalar distributions in a different way.  In the case of spherical symmetry, although the perfect fluid analogy is still allowed (we will discuss that point later in Section~\ref{sec.discussion}), a static, radial-dependent scalar field $\phi=\phi(r)$ 
is required in most physical situations. For a static field the derivative terms are space-like, $X<0$, and the anisotropic description necessary. This suggests the following attractive picture:  DM could mimic a perfect fluid ``dust'' in cosmology, but a(n anisotropic) relativistic one in galaxies. This is not possible for a standard perfect fluid, where the observed non-relativistic rotation curves guarantee a Newtonian description, $p\ll\rho$; see the Appendix~\ref{app0}. (Do not confuse the Newtonian with the weak-field regime.) This route has been followed by many authors before~\cite{Matos:2000ki, ArmendarizPicon:2005nz, matos2}, however, as we find next, there are some crucial aspects of these configurations that have been overlooked until now.

For a static, spherically symmetric configuration, the most general expression for the spacetime metric (in polar-areal coordinates, such that spheres of constant $r$ have area $4\pi r^2$) takes the form
\begin{equation}
ds^{2}=-\exp(2\psi(r)) dt^{2}+h(r)dr^{2}+r^{2}d\Omega\,. \label{line}
\end{equation}
The effective gravitational potential $\psi(r)$ and the metric function $h(r)>0$ are dimensionless, and $d\Omega = d\theta^2 + \sin^2 d\varphi^2$ is the standard solid angle element in three dimensions, with $r\in[0,\infty)$. A regular spacetime metric demands $\psi(r= 0)=\textrm{const.}$, $h(r= 0)=1$; attractive spacetime configurations $d \psi/dr \ge 0$. Note that Eq.~(\ref{line}) is only a possible parametrization for the spacetime metric, and it does not contain any physical content beyond the underlying symmetries. 

The most general expression for the energy-momentum tensor compatible with the spacetime symmetries is given by
\begin{equation}
T_{\mu\nu}=(\rho+p_{\perp})u_{\mu}u_{\nu}+p_{\perp}g_{\mu\nu}+(p_{\|}-p_{\perp})n_{\mu}n_{\nu}\,.\label{energymomentum}
\end{equation}
Here $\rho$ is the energy density, $p_{\|}$ the pressure in the direction parallel to $n_{\mu}$, and $p_{\perp}$ the pressure in an orthogonal direction, all measured by an observer at rest with respect to the four-velocity $u_{\mu}$. For static, spherically symmetric configurations $u_{\mu}=(-\exp(\psi),0,0,0)$, and $n_{\mu}=(0,h^{1/2},0,0)$. Regularity at the origin demands $\rho(r=0)=\rho_0$,
$p_{\|}(r=0)=p_{\|0}$, and $p_{\perp}(r=0)=p_{\perp 0}$, with $\rho_0$, $p_{\|0}$ and $p_{\perp 0}$ all finite. By localized matter distribution we shall mean one where $\rho(r\to\infty)=p_{\|}(r\to\infty)=p_{\perp}(r\to\infty)=0$.

\subsection{A first proof of the no-go theorem}\label{proof1}

Now we can prove the main result of this paper:  that a static scalar field $\phi=\phi(r)$ can source no static, spherically symmetric, regular, spatially localized, attractive, stable spacetime configuration with positive definite energy density.  We use the Appendix~\ref{app1} to show that, for the canonical and the purely-kinetic scalar fields, these configurations are not possible
even at the expense of having negative energy densities.

The argument is simple, and it relies on the impossibility of fulfilling all the previous conditions at the same time. From the energy-momentum conservation, $\nabla_{\mu}T^{\mu\nu}=0$, we obtain the equation for hydrostatic equilibrium,
\begin{equation}
 \frac{dp_{\|}}{dr}=-(\rho+p_{\|})\,\frac{d \psi}{dr} - \frac{2(p_{\|}-p_{\perp})}{r}\, . \label{hydrostatic-equilibrium}
\end{equation}
For the case of a static scalar field, $X<0$, the identities in Eq.~(\ref{relani}) lead to $\rho+p_{\|}=p_{\|}-p_{\perp}=-2X \partial\mathcal{L}/\partial X$. In order to avoid tachyons and ghosts, we should satisfy $\partial\mathcal{L}/\partial X >0$; see the Appendix~\ref{app2} for details. Then a static, spherically symmetric, stable, attractive spacetime sourced by a static scalar field requires $dp_{\|}/dr < 0$ for hydrostatic equilibrium. This condition, together with that for a localized matter distribution, $p_{\|}(r\to\infty)=0$, guarantees a positive definite radial pressure, $p_{\|}(r)>0$. A regular spacetime metric demands $\partial_r\phi(r= 0)=0$, i.e. $X(r= 0)=0$, and then, $\rho_{0}=-p_{\| 0}=-p_{\perp 0}$. In particular, since $p_{\|}$ is positive definite, that implies $\rho_0 < 0$, i.e. the energy density should be negative, at least close to the center of the configuration. 
 
In general relativity not only the energy density but the combination $\rho+p_{\perp}+2p_{\|}$ sources gravity~\cite{ehlers}. For static scalar fields $\rho_0+p_{\perp 0}+2p_{\| 0} = -2\rho_0$, and then it is natural to understand why attractive spacetime configurations with positive energy density are not possible in general relativity. Note, however, that Eq.~(\ref{hydrostatic-equilibrium}) and the paragraph below are generic, and apply for any metric theory, i.e. gravity is described by the metric tensor of a spacetime manifold, with test particles following timelike geodesics. In the weak-field regime, $r |d\psi/dr|\ll 1$, the last term in Eq.~(\ref{hydrostatic-equilibrium}) dominates, and it is not necessary to assume an attractive gravitational potential.

It is pertinent to mention a couple of examples where the theorem holds. Negative energy densities are present in the analytic solution reported in\footnote{A close inspection of Eq.~(19) reveals a negative definite energy density. [There is a typo in the printed version, where the right-hand side of Eq.~(19) should be positive definite; see astro-ph/0003398.]} Ref.~\cite{Matos:2000ki}, where the authors demand halos with flat rotation curves, and also in the numerical solutions obtained in Ref.~\cite{matos2}, where the condition on the rotational curves is relaxed. (See also Ref.~\cite{kodama} for a previous discussion of static scalar field configurations in the strong-field regime.) If the scalar fields are non-canonical, see Refs.~\cite{ArmendarizPicon:2005nz}. Here we show that negative energy densities are generic, and they are not restricted to the particular solutions in Refs.~\cite{Matos:2000ki,ArmendarizPicon:2005nz,matos2,kodama}.

As applied to the galaxies in the Universe, this no-go theorem assumes a very simple model for the galactic halos. One could probably argue that the presence of baryons might play an important role in a more realistic model, particularly close to the center of these configurations, where the negative energy densities were identified. We do not expect to recover all the physical properties of the halo without taking into account the existence of  other matter sources in galaxies, but we consider baryonic matter cannot be an essential ingredient for the main existence of these configurations. After all, according to the standard cosmological picture, DM sourced the primordial wells for the subsequent development of cosmic structure. Furthermore, we know of the existence of dwarf galaxies which are DM dominated~\cite{dwarf1}. Even so and for the more skeptical of our readers  we use some lines to show that the presence of additional matter sources cannot avoid the appearance of negative energy densities.

\subsection{A second proof of the no-go theorem}\label{sec.second}

As was noted in Refs.~\cite{Matos:2000ki, Nucamendi:2000jw}, we can always write the effective gravitational potential in the form
\begin{equation}
 \psi(r) = \int_r^{\infty} \frac{v_c^2(r)}{r} dr \, , \label{para.B}
\end{equation}
with $0\le v_c(r)<1$ the velocities of the test particles in circular motion around $r=0$. A regular spacetime metric satisfies $v_c^2(r=0)=0$; an attractive gravitational potential $v_c^2(r)\ge 0$.

Introducing Eqs.~(\ref{line}) and (\ref{energymomentum}) into Einstein equations, and using the expression in Eq.~(\ref{para.B}), we get
\begin{subequations}\label{eins}
\begin{eqnarray}
\frac{1}{hr^{2}}\left[\frac{h'}{h}r+h-1\right]&=& 2\rho\,, \label{eins1} \\ 
\frac{1}{hr^{2}}\left[(1+\ell)-h\right]&=& 2p_{\|}\,, \label{eins2} \\
-\frac{1}{4hr^{2}}\left[(2+\ell)\frac{h'}{h}r-(\ell^2+2r\ell')\right]&=& 2p_{\perp}\,.  \label{eins3}
\end{eqnarray}
\end{subequations}
The prime here denotes the derivative with respect to the radial coordinate, and we have introduced $\ell(r)=2v_c^2(r)$. Eqs.~(\ref{eins1}) and (\ref{eins3}) can be combined to obtain
\begin{equation}\label{relation.previa}
(\ell+2)\rho + 4p_{\perp} = \left[\frac{(2+\ell)m}{r}+\frac{\ell^{2}+2r\ell'}{2h}\right]\frac{1}{r^2}\,.
\end{equation}
As usual, the effective gravitational mass $m$ is defined from $h= 1/(1-2m/r)$, with $m(r)= \int^{r}_{0} \rho(r)r^{2} dr$.

Eq.~(\ref{relation.previa}) is valid for all values of the radial coordinate, and for all the static, spherically symmetric configurations. In galaxies, baryons and DM contribute to $\rho$ and $p_{\perp}$. However, for those regions dominated by a static scalar field, if any, $p_{\perp}=-\rho$, Eq.~(\ref{relation.previa}) simplifies to
\begin{equation}
\rho=-\frac{1}{2-\ell}\left[\frac{(2+\ell)m}{r}+\frac{\ell^{2}+2r\ell'}{2h}\right]\frac{1}{r^2}\,.\label{relation}
\end{equation}
As long as $\ell^2+2r\ell'> 0$, the only possible way to have a positive energy density is with a negative effective gravitational mass; and the opposite, a positive effective gravitational mass requires a negative energy density. A static scalar field with positive energy density still seems possible, but with a negative effective gravitational mass. However, both conditions are not compatible with a regular spacetime metric: in order to have $m<0$ for some value of the radial coordinate, say $r=r_0$, we should demand $\rho<0$, at least for some region in the interval $0<r<r_0$. 

For regular, attractive spacetime configurations, $\ell(r=0)=0$, $\ell(r>0)>0$, the condition $\ell^2+2r\ell'> 0$ is guaranteed for some region in the distribution. If we restrict our attention to idealized galaxies without baryons, Eq.~(\ref{relation}) is satisfied for all values of the radial coordinate, recovering the negative energy densities we identified close to the center of the configuration by means of the previous argument in Section~\ref{proof1}.

In a more realistic model, one should consider the presence of other matter sources. In spiral galaxies, for instance, baryons and DM can contribute equally to the mass within the optical radius~\cite{Williams:2009zm}. However, the external regions of spiral galaxies 
where (nearly) flat rotation curves are observed are dominated by DM. There are several examples of galaxies for which the relation $\ell^2+2r\ell'> 0$ is still valid in the outer regions, see for instance NGC~2403 and NGC~3621 in Ref.~\cite{deBlok:2008wp}. Again, the negative energy densities emerge, no matter what you could have in the core of the galaxy. Note that, contrary to the first proof in Section~\ref{proof1}, we used Einstein equations, but this time it was not necessary to assume a stable theory.

\section{Discussion}\label{sec.discussion}

All these results extend trivially to the case with more than one field in the action. For a generic theory a set of static scalars $\phi_i=\phi_i(r)$ is necessary in order to  recover a static spacetime background.  Here $i=1,\ldots, n$ labels the different fields. However, if there is an internal continuous global symmetry, $\phi_i\to\phi_i'=\mathcal{T}(\alpha,\phi_j)$, the staticity of the spacetime metric can be recovered in another way: by proposing a solution of the form $\phi_i(t,r)=\mathcal{T}(\omega t,\varphi_j(r))$. Here $\alpha$ is a continuous parameter for the transformation, and $\omega$ a constant with dimensions of the inverse of time. The fields are now dynamical, but the spacetime metric is static. 

A case of particular interest is that of a canonical scalar field with an internal $U(1)$ symmetry, $\phi\to e^{i\alpha}\phi$,
leading to boson stars~\cite{Ruffini:1969qy}: scalar field configurations of the form $\phi(t,r)=e^{i\omega t}\varphi(r)$. Now, for specific radial functions $\varphi(r)$ and particular values of the constant $\omega$, there are static, spherically symmetric, regular, spatially localized, attractive, stable, self-gravitating configurations, but at the expense of a non-zero Noether charge in the system: the difference between the number of particles and antiparticles. This charge is associated with the time-dependency of the scalar field, and then the arguments in Section~\ref{sec.static} do not apply.

Another example of interest is provided by perfect fluids. The action principle describing a perfect fluid in general relativity can be written in terms of the velocity potentials~\cite{Schutz}. Spherical symmetry guarantees no vorticity, and then $u_{\mu}=\partial_{\mu}\varphi$, with $u_{\mu}$ the four-velocity of the fluid. The action for a perfect fluid is invariant under shift-transformations in the velocity potential, $\varphi\to \varphi+\textrm{const.}$, and it is this symmetry that makes possible the existence of static, spherically symmetric, regular, spatially localized, attractive, stable perfect fluid configurations with positive definite energy density~\cite{Tolman:1939jz}; see also Ref.~\cite{DiezTejedor:2006qh}. In this case the conserved Noether charge associated to the shift-invariance is the total entropy in the system~\cite{Diez-Tejedor:2013nwa}.

As with any no-go theorem, the results in this paper can be circumvented by relaxing some of the initial assumptions: possible ways out involve dynamical spacetimes~\cite{Seidel:1991zh}, galactic halos made of smaller mini-halos~\cite{Barranco:2010ib}, thermal distributions for the scalar field~\cite{Bilic:2000ef}, and dark sectors with more fields and (gauge) symmetries~\cite{evslin}.

\begin{acknowledgments}
We are grateful to Jarah Evslin, Alex Feinstein, Robert Scherrer, and Luis Ure\~na-Lopez for useful comments and discussions. This work was partially supported by PIFI, PROMEP, DAIP-UG, CAIP-UG, the ``Instituto Avanzado de Cosmolog\'ia'' (IAC) collaboration, DGAPA-UNAM Grant No. IN115311, and CONACyT M\'exico under Grants No. 182445 and No. 167335.
\end{acknowledgments}


\appendix

\section{Perfect fluid halos are Newtonian}\label{app0}

The three conditions below, when satisfied simultaneously, guaranty the viability of a Newtonian description: $i)$ Weak gravitational fields, $g_{\mu\nu}=\eta_{\mu\nu}+\gamma_{\mu\nu}$, 
$ii)$ Negligible stresses when compared to the mass-energy density, $p_{ij}\ll\rho$, and, $iii)$ Relative motions much smaller than the speed of light, $u^i\ll 1$. (See for instance Ref.~\cite{wald} for details.) Here $\eta_{\mu\nu}=\textrm{diag}(-1,1,1,1)$ is the Minkowski spacetime metric, $\gamma_{\mu\nu}\ll 1$ a measure for the deviations with respect to the Minkowski metric, $p_{ij}$ the spatial stresses, and $u^{\mu}=(u^0,u^i)$ the four-velocity for the particles in the configuration.

Perfect fluids satisfy $p_{\|}=p_{\perp}$. Combining this identity with Eqs.~(\ref{eins2}) and (\ref{eins3}), we obtain a differential equation for the metric function~$h(r)$,
\begin{equation}\label{eq.dif}
 (2+\ell)\frac{h'}{h}r-4h+(4+4\ell-2r\ell'-\ell^2)=0 \, .
\end{equation}
Astrophysical observations provide $\ell(r)\lesssim 10^{-5}$. To the first order in $\ell$, the solution to the Eq.~(\ref{eq.dif}) that is regular at the origin takes the form $h(r)=1+\ell(r)$.
Introducing this expression into the Eqs.~(\ref{eins}), we can read $u^{i}_{\textrm{test}}\sim \mathcal{O}(\ell^{1/2})$,  $\gamma,\, \rho \sim\mathcal{O}(\ell)$,  $p\sim\mathcal{O}(\ell^2)$, and $u^i_{\textrm{halo}}=0$, i.e. perfect fluid halos are Newtonian objects. This is no longer true for the static scalar field configurations, where $p_{\perp}=-\rho$; see Eqs.~(\ref{relani}) above.

\section{The canonical and the purely-kinetic scalar fields}\label{app1} 

Localized, regular canonical scalar field configurations satisfy $\partial_r\phi(r= 0)=0$, $\phi(r\to\infty)=\textrm{const.}$. Together with the Klein-Gordon equation, $\Box \phi - \partial\mathcal{L}/\partial\phi =0$, this implies $\partial\mathcal{L}/\partial\phi (r=0) = \partial\mathcal{L}/\partial\phi (r\to\infty) = 0$. That is, for two different values of the scalar field,  $\phi(r=0)=\phi_0$ and $\phi(r\to\infty)=\phi_{\infty}$, we need to satisfy $\partial\mathcal{L}/\partial\phi |_{\phi_0} = 
\partial\mathcal{L}/\partial\phi |_{\phi_\infty} = 0$. This is possible only if $\partial^2\mathcal{L}/\partial\phi^2$ changes sign between $\phi_0$ and $\phi_{\infty}$,  signaling the appearance of tachyons in the low-energy spectra; see Eq.~(\ref{no-instabilities.simple}) in the Appendix~\ref{app2}. Here it has not been necessary to assume an attractive effective gravitational potential.

For the case of a purely-kinetic scalar field, deriving Eq.~(\ref{relani}) for $p_{\|}$ 
with respect to the radial coordinate, we obtain
\begin{equation}
 \frac{d p_{\|}}{dr}= -\left(\frac{\partial\mathcal{L}}{\partial X}+2X\frac{\partial^2 \mathcal{L}}{\partial X^2}\right)\frac{\partial X}{\partial r}\,.\label{proof}
\end{equation}
Regular, static scalar field configurations satisfy $X(r=0)=0$, $X(r>0)\le 0$. That is, the sign of $\partial X/\partial r$ is negative, at least for some values of the radial coordinate.
Since $d p_{\|}/dr<0$ for hydrostatic equilibrium, the sign of $\partial\mathcal{L}/\partial X+2X\partial^2 \mathcal{L}/\partial X^2$ should be negative also, at least for this same interval with negative gradients of the kinetic scalar, signaling the appearance of tachyons in the low-energy spectra;  see again Eq.~(\ref{no-instabilities.simple}) in the Appendix~\ref{app2}.

\section{Absence of tachyons and ghosts in the low-energy spectra}\label{app2}

In order to have a sensible theory, at least at the effective level, we should avoid the appearance of classical and quantum instabilities in the spectrum of low-energy perturbations.

Let us consider the behavior of the small perturbations around a static,  spherically symmetric scalar field configuration. Two comments are in order here. First, we will consider only perturbations in the scalar field, neglecting any possible backreaction on the metric tensor. Second, since any regular spacetime metric is locally Minkowski, we can restrict our analysis to flat spacetime.  We can then propose a solution of the form  $\phi(t,\vec{x})=\phi_0(z)+\delta\phi(t,\vec{x})$, with  $\phi_0(z)$ the background solution and $z$ signaling the direction of the field gradients, $\vec{x}=(x,y,z)$. Expanding Eq.~(\ref{action}) to the quadratic order in field perturbations, $\delta\phi(t,\vec{x})$, we obtain
\begin{subequations}
\begin{eqnarray}
 \mathcal{L}\sim c_1(\partial_0\delta\phi)^2-c_1(\partial_{\perp}\delta\phi)^2 - c_2(\partial_z\delta\phi)^2-
\hspace{.7cm}&&\nonumber\\
&& \hspace{-3.95cm} 
2c_3 (\partial_z\delta\phi)\delta\phi - c_4 (\delta\phi)^2 \, ,
\end{eqnarray}
with $(\partial_\perp \delta\phi)^2= (\partial_x\delta\phi)^2+(\partial_y\delta\phi)^2$, and
\begin{eqnarray}
&&c_1=\frac{\partial\mathcal{L}}{\partial X}\,, \quad
c_2=\frac{\partial\mathcal{L}}{\partial X}+2X\frac{\partial^2\mathcal{L}}{\partial X^2}\,, \\
&&c_3=\frac{1}{2} \frac{\partial^2 \mathcal{L}}{\partial X \partial \phi}\partial_{z}\phi  \,, \quad
\; c_4=-\frac{\partial^2\mathcal{L}}{\partial\phi^2}\, .
\end{eqnarray}
\end{subequations}
All these quantities are evaluated at $\phi_0$, $2 X_0=-(\partial_z\phi_0)^2$. In order to have a positive definite Hamiltonian density, we should satisfy
\begin{equation}\label{conditions}
c_1>0\,, \quad c_{+}\pm \delta \ge 0\,,
\end{equation}
where $c_{\pm}=(c_2\pm c_4)/2$, and $\delta^2=c^2_{-}+c_3^2$. All the conditions in Eq.~(\ref{conditions}) are necessary in order to avoid tachyons.  The first condition, $\partial\mathcal{L}/\partial X >0$, guarantees the absence of ghosts. [Here we are only proving the local (in)stability of the configurations, but global considerations could make them stable, e.g. global monopoles~\cite{Barriola:1989hx}.]

For the particular case in which $c_3=0$ (a canonical scalar field, for instance, or a purely-kinetic theory), the~conditions
\begin{equation}\label{no-instabilities.simple}
 \frac{\partial\mathcal{L}}{\partial X} > 0\,,\quad
 \frac{\partial\mathcal{L}}{\partial X}+2X\frac{\partial^2\mathcal{L}}{\partial X^2} \ge 0\,,\quad
 \frac{\partial^2\mathcal{L}}{\partial \phi^2} \le 0\,, 
\end{equation}
guarantee the absence of classical and quantum instabilities. Notice that these conditions coincide with those obtained for the  stability of a homogeneous and isotropic background.

\end{document}